\documentclass[seceq]{ptptex}

\title{Thermodynamics of Apparent Horizon in Brane World Scenarios }

\author{Rong-Gen Cai~\footnote{Email: cairg@itp.ac.cn}}

\inst{Kavli Institute for Theoretical Physics China(KITPC)\\
      at the Chinese Academy of Sciences \\
      Institute of Theoretical Physics, Chinese Academy of
Sciences, P.O. Box 2735, Beijing 100080, China }

\abst{Applying the Clausius relation, $\delta Q=TdS$, to the
apparent horizon of FRW universe in brane world scenarios, we show
that an explicit entropy expression associated with the apparent
horizon can be obtained. On the apparent horizon, the relation,
$dE=TdS +WdV$, also holds in the brane world scenarios. We show
these results in the RSII model, warped DGP model and the more
general case with a Gauss-Bonnet term in the bulk and an intrinsic
curvature term on the brane. {\bf CAS-KITPC/ITP-027}}

\begin{document}

\maketitle

\section{Introduction}

General relativity together with properties of black hole horizon
leads to four laws of black hole mechanics~\cite{BCH}, which are
quite similar to four laws of thermodynamics. Due to Hawking's
discovery that black hole radiates particles with thermal
spectrum~\cite{Hawking}, it turns out that it is not just analogy
between the four laws of black hole mechanics and four laws of
thermodynamics. Black hole is nothing, but an ordinary thermal
object. However, as a thermal object, black hole is special in the
sense that black hole has a temperature proportional to its
surface gravity on the horizon and an entropy proportional to its
horizon area~\cite{Bek}. Since both the surface gravity and
horizon area are geometric quantities, it seemingly implies that
there exists some relation between Einstein equations, describing
dynamics of spacetime, and thermodynamic laws, describing some
relations among thermodynamic quantities.

Jacobson is the first one to seriously investigate such a
relation~\cite{Jaco}. Jacobson finds that it is indeed possible to
derive the Einstein  equations from the proportionality of entropy
to the horizon area together with the fundamental relation,
$\delta Q= TdS$, the Clausius relation in thermodynamics, assuming
the relation holds for all local Rindler causal horizons through
each spacetime point. Here $\delta Q$ and $T$ are the energy flux
and Unruh temperature seen by an accelerated observer just inside
the horizon. In the cosmological set up with a
Friedmann-Robertson-Walker (FRW) metric, applying the Clausius
relation together with the continuity relation to the apparent
horizon of the FRW universe, one is able to derive the Friedmann
equations, describing the dynamical evolution of the
universe~\cite{FK,Dan,Bous,CK}. In particular, with the relation
between the horizon entropy and horizon geometry of black holes in
Lovelock gravity, the same approach reproduces the corresponding
Friedmann equations of FRW universe in the Lovelook
gravity~\cite{CK}. Naively applying the approach to the
scalar-tensor theory and $f(R)$ theory, one fails to reproduce
corresponding Friedmann equations in those theories~\cite{AC1}.
Jacobson {\it et al} argue that for $f(R)$ gravity, in order to
derive the field equations, a non-equilibrium thermodynamics set
up should be employed, where an entropy production term should be
added to the Clausius relation~\cite{Jacob2,AC3}. This statement
also holds in the scalar-tensor gravity~\cite{CC1}. Furthermore,
we have shown that at the apparent horizon of the FRW universe,
Friedmann equations can be rewritten in a universal form $dE= TdS
+WdV$, not only in Einstein's general relativity, but also in
Lovelock gravity~\cite{AC2,CC1}, here $E$ is the energy of matters
inside the apparent horizon, $V$ is the volume of the space
surrounded by the apparent horizon, $W=(\rho-p)/2$ with $\rho$ and
$p$ being the energy density and pressure of matters in the
universe, and $T$ and $S$ are temperature and entropy associated
with the apparent horizon. In Einstein's general relativity, in
fact, the universal relation can be understood as the unified
first law first proposed by Hayward~\cite{Hawy}.

Over the past years a lot of attention has been attracted to the
brane world scenario. The so-called brane world scenario assumes
that our universe is a 3-dimensional brane embedded a higher
dimensional bulk spacetime; the matters in particle physical
standard model are confined on the brane, while gravity can
propagate in the whole bulk spacetime~\cite{ADD,RS}. Due to the
existence of the brane, the gravity on the brane is no longer the
Einstein's general relativity. As a result,  it is interesting to
apply this method developed in \cite{CC1} to study the entropy of
the apparent horizon of a FRW universe in the brane world
scenario. This is partially because the gravity on the brane is
not the Einstein theory, the well-known area formula for black
hole horizon entropy must not hold in this case, and partially
because exact analytic black hole solutions on the brane have not
been found so far, and then it is not known how the horizon
entropy of black hole on the brane is determined by the horizon
geometry. On the other hand, the exact Friedmann equations of the
FRW universe on the brane have been derived for the RSII model
some years ago~\cite{Langlois}.

In this paper we summarize the results obtained
in~\cite{CC2,SWC1,SWC2}. It is found that in the various brane
world scenarios, not only does the universal relation $dE=TdS
+WdV$ hold at the apparent horizon of FRW universe on the brane,
but also one can get an entropy expression in terms of horizon
geometry associated with the apparent horizon. In the next
section, we will discuss the thermodynamics of apparent horizon in
the RSII model. The case of warped DGP model will be studied in
Sec.~3. The more general case with a Gauss-Bonnet term in the bulk
and an intrinsic curvature term on the brane will be investigated
in Sec.~4.

Here we also would like to mention some recent discussions on the
relation between the Freidmann equation and the first
law~\cite{Ge,GW,Wu1}. In addition, Padmanabhan and his
collaborators studied the relation between the Einstein equations
and the first law of thermodynamics in the setting of black hole
spacetime~\cite{Pad}.

\section{ Thermodynamics of apparent horizon in RSII model}

Hayward has proposed a method to deal with thermodynamics
associated with trapping horizon of a dynamical black hole in
$4$-dimensional Einstein theory~\cite{Hayward2,Hawy}. In this
method, for spherical symmetric space-times, $ds^2
=h_{\alpha\beta}dx^{\alpha} dx^{\beta} + r^2 d\Omega_2^2$,
Einstein equations can be rewritten in a form called ``unified
first law"
\begin{equation}
\label{2eq1}
 dE=A\Psi+WdV\, ,
\end{equation}
where $E$ is the so-called Misner-Sharp energy, defined by
$E=\frac{r}{2G}(1-h^{\alpha\beta}\partial_{\alpha}r
\partial_{\beta}r)$; $A=4\pi r^2$ is the sphere area with radius $r$ and
$V=4\pi r^3/3$ is the volume; and the work density $W \equiv
-\frac{1}{2}T^{\alpha\beta}h_{\alpha\beta}$ and the energy supply
vector $\Psi_{\alpha} \equiv T_{\alpha}^{\ \beta}\partial_{\beta}r
+W
\partial_{\alpha}r$ with $T_{\alpha\beta}$ being the
energy-momentum tensor of matter in the spacetime.
 Projecting this unified first law along a trapping
horizon, one gets the first law of thermodynamics for dynamical
black hole
\begin{equation}
\label{2eq2}
 \langle dE,\xi\rangle=\frac{\kappa}{8\pi G}\langle
dA,\xi\rangle+W\langle dV,\xi\rangle\, ,
\end{equation}
where  $\kappa$ is surface gravity, defined by $\kappa
=\frac{1}{2\sqrt{-h}}\partial_{\alpha}(\sqrt{-h}h^{\alpha\beta}
\partial_{\beta}r)$,
on the trapping horizon, $\xi$ is a projecting vector. In the
paper~\cite{CC1}, we have applied this theory to study the
thermodynamics of apparent horizon of a FRW universe in higher
dimensional Einstein gravity and some non-Einstein theories, such
as Lovelock gravity and scalar-tensor gravity by rewriting gravity
field equations to standard Einstein equations with a total
energy-momentum tensor.  The total energy momentum tensor consists
of two parts: one is just the ordinary matter energy-momentum
tensor; and the other is an effective one coming from the
contribution of the higher derivative terms (for example, in
Lovelock gravity) and scalar field (for example, in scalar-tensor
gravity theory). The total energy-momentum tensor will enter into
the energy-supply $\Psi$ and work term $W$ in (\ref{2eq1}).
Therefore, the work density and energy-supply vector in
(\ref{2eq1}) also can be decomposed into ordinary matter part and
effective part: $\Psi=\stackrel{m}{\Psi}+\stackrel{e}{\Psi}$,
$W=\stackrel{m}{W}+\stackrel{e}{W}$. That is, $
\stackrel{m}{\Psi}$ and $\stackrel{m}{W}$ are the energy-supply
vector and work density from the ordinary matter contribution,
while $\stackrel{e}{\Psi}$ and $\stackrel{e}{W}$ comes from the
effective energy-momentum tensor. The matter part of energy-supply
is the energy flux defined by pure ordinary matter, so its
integration on the sphere (after projecting along the apparent
horizon) naturally defines the heat flow $\delta Q $ in the
Clausius relation $\delta Q=T dS$.  On the other hand, the unified
first law tells us
\begin{equation}
\label{2eq3}
 \delta Q=\langle A\stackrel{m}{\Psi}
,\xi\rangle=\frac{\kappa}{8\pi G}\langle dA,\xi\rangle-\langle
A\stackrel{e}{\Psi} ,\xi\rangle\, .
\end{equation}
The right hand side of the above equation should be written in the
form of $T\delta S$ by considering spacetime horizon with
$T=\kappa/2\pi$ as an  equilibrium thermodynamic system. Thus we
could get an entropy expression because the right hand side of the
above equation is easily calculated. In this section, we apply
this approach to the case of RSII model.

The so-called RSII model is a 3-brane embedded in a 5-dimensional
anti-de Sitter (AdS) space~\cite{RS}. To be more general, we are
considering an $(n-1)$-brane with positive tension in an
$(n+1)$-dimensional AdS space. Following the method given in
\cite{SMS}, we get the effective gravity field equations on the
brane with tension $\lambda$:
\begin{eqnarray}
{}^{(n)}G_{\mu\nu}=-\Lambda_n q_{\mu\nu} + 8 \pi
G_n\tau_{\mu\nu}+\kappa_{n+1}^4\,\pi_{\mu\nu} -E_{\mu\nu}\,,
\label{2eq4}
\end{eqnarray}
where
\begin{equation}
G_n=\frac{n-2}{32\pi(n-1)}\lambda\kappa_{n+1}^4\,, \label{GNdef}
\end{equation}
\begin{equation}
\Lambda_n=\kappa_{n+1}^2 \left(\frac{n-2}{n}\Lambda_{n+1}
+\frac{n-2}{8(n-1)}\kappa_{n+1}^2\,\lambda^2\right)\,,
\label{2eq6}
\end{equation}
\begin{equation}
\pi_{\mu\nu}= -\frac{1}{4} \tau_{\mu\alpha}\tau_\nu^{~\alpha}
+\frac{1}{4(n-1)}\tau\tau_{\mu\nu}
+\frac{1}{8}q_{\mu\nu}\tau_{\alpha\beta}\tau^{\alpha\beta}-\frac{1}{8(n-1)}
q_{\mu\nu}\tau^2\,, \label{2eq7}
\end{equation}
and $E_{\mu\nu}$ is the electric part of the $(n+1)$-dimensional
Weyl tensor. We now restrict ourself to the case with vanishing
cosmological constant on the brane, which gives the constraint
\begin{equation}
\label{2eq8}
 \Lambda_{n+1}=-\frac{n(n-1)}{2\kappa_{n+1}^2\ell^2}\,
,\quad \lambda=\frac{2(n-1)}{ \kappa_{n+1}^2\ell}\, .
\end{equation}
 This leads to the relation between the gravity constants on the
 brane and in the bulk
\begin{equation}
\label{2eq9}
 \frac{G_{n+1}}{G_{n}}=\frac{ \kappa_{n+1}^2}{8\pi
 G_{n}}=\frac{2\ell}{n-2}\, .
\end{equation}
Considering a perfect fluid with energy-stress tensor
\begin{equation}
\label{2eq10} \tau_{\mu\nu}= (\rho+p) t_{\mu}t_{\nu}+pq_{\mu\nu},
\end{equation}
on the brane, the gravity field equations on the brane reduce to
\begin{eqnarray}
 {}^{(n)}G_{\mu\nu}&=& 8 \pi
G_n\left[\rho+p+\frac{n-2}{n-1}\frac{\kappa_{n+1}^4}{32 \pi
G_n}\rho(\rho+p)\right]t_{\mu}t_{\nu}\nonumber \\
&&+ 8 \pi G_n\left[p+\frac{n-2}{n-1}\frac{\kappa_{n+1}^4}{64 \pi
G_n}\rho(\rho+2p)\right]q_{\mu\nu}-E_{\mu\nu} \, .
\end{eqnarray}
Further, we set the bulk to be a pure AdS space so that the Weyl
tensor vanishes, and the metric on the brane to be an
$n$-dimensional FRW metric
\begin{eqnarray}
\label{2eq12}
ds^2&=&q_{\mu\nu}dx^{\mu}dx^{\nu}=-dt^2+\frac{a(t)^2}{1-kr^2}dr^2+a(t)^2r^2d\Omega_{n-2}^2
\nonumber\\
&=&\tilde{h}_{ab}dx^a dx^b+\tilde{r}^2d\Omega_{n-2}^2\, ,
\end{eqnarray}
where $\tilde{r}=a(t)r$, $x^0=t, x^1=r$. Thus we can obtain the
Friedmann equations on the brane without dark radiation term
($E_{\mu\nu}=0$)
\begin{equation}
\label{2e13}
 (n-1)(n-2)\left(H^2+\frac{k}{a^2}\right)=16\pi G_n\left(
\rho+\frac{n-2}{n-1}\frac{\kappa_{n+1}^4}{64 \pi
G_n}\rho^2\right)\, ,
\end{equation}
\begin{equation}
\label{2e14}
 -(n-2)\left(\dot{H}-\frac{k}{a^2}\right) =8\pi G_n
\left(\rho+p+\frac{n-2}{n-1}\frac{\kappa_{n+1}^4}{32 \pi
G_n}\rho(\rho+p) \right)\, .
\end{equation}
The energy density $\rho$ on the brane satisfies the continuity
equation
\begin{equation}
\label{2e15}
 \dot{\rho}+(n-1)H(\rho+p)=0\, .
\end{equation}
This equation and the Friedmann equations will be used later.

By introducing an effective stress-energy tensor, one can rewrite
the gravitational field equations on the brane to
\begin{equation}
\label{2eq16}
 {}^{(n)}G_{\mu\nu}= 8 \pi
G_n\left(\tau_{\mu\nu}+\stackrel{e}{\tau}_{\mu\nu}\right)\, ,
\end{equation}
where
\begin{equation}
\stackrel{e}{\tau}_{\mu\nu}=\frac{\kappa_{n+1}^4}{8 \pi
G_n}\pi_{\mu\nu}\, .
\end{equation}
Equations (\ref{2eq16}) are in the standard form of Einstein
equations, so the unified first law is applicable to that case.
Thus the work density term  can have the following form
\begin{equation}
W=\stackrel{m}{W}+\stackrel{e}{W},
\end{equation}
with
\begin{equation}
\stackrel{m}{W}=\frac{1}{2}(\rho-p)\, ,\quad\quad
\stackrel{e}{W}=-\frac{n-2}{n-1}\frac{\kappa_{n+1}^4}{64\pi G_{n}}
\rho p\, ,
\end{equation}
and similarly, the energy supply vector can be decomposed as
\begin{equation}
\Psi=\stackrel{m}{\Psi}+\stackrel{e}{\Psi},
\end{equation}
with
\begin{equation}
\stackrel{m}{\Psi}=-\frac{1}{2}(\rho+p)H\tilde{r}dt+\frac{1}{2}(\rho+p)adr\,
,
\end{equation}
\begin{equation}
\stackrel{e}{\Psi}=-\frac{n-2}{n-1}\frac{\kappa_{n+1}^4}{64\pi
G_{n}}\rho(\rho+p)H\tilde{r}dt+\frac{n-2}{n-1}\frac{\kappa_{n+1}^4}{64\pi
G_{n}}\rho(\rho+p)adr\,.
\end{equation}

Using these quantities and the Misner-Sharp energy in $n$
dimensions inside the apparent horizon
\begin{equation}
E=\frac{1}{16\pi G_n}(n-2)\Omega_{n-2}\tilde{r}_A^{n-3}\, ,
\end{equation}
where $A_{n-2}=\Omega_{n-2}\tilde r_A^{n-2}$ and $V_{n-2}=A_{n-2}
\tilde{r}_A/(n-1)$ are the area and volume of the $(n-2)$- sphere
with the apparent horizon radius $\tilde{r}_A=1/\sqrt{H^2+k/a^2}$
of the FRW universe, respectively, we can put the $(00)$ component
of equations of motion (\ref{2eq16})  into the form of the unified
first law
\begin{equation}
dE=A_{n-2}\Psi+WdV_{n-2}\, .
\end{equation}
After projecting along a vector
$\xi=\partial_t-(1-2\epsilon)Hr\partial_r$ with
$\epsilon=\dot{\tilde{r}}_A/2H\tilde{r}_A$, we get the first law
of thermodynamics of the apparent horizon~\cite{CC1}
\begin{equation}
\langle dE, \xi\rangle=\frac{\kappa}{8\pi G_n}\langle dA_{n-2},
\xi\rangle+\langle WdV_{n-2}, \xi\rangle\,,
\end{equation}
where $\kappa=-(1-\dot {\tilde r}_A/(2H\tilde r_A))/\tilde r_A$ is
the surface gravity of the apparent horizon. The pure matter
energy-supply $A_{n-2}\stackrel{m}{\Psi}$ (after projecting along
the apparent horizon) gives the heat flow $\delta Q$ in the
Clausius relation $\delta Q=T d S$. By using the unified first law
on the apparent horizon, we have
\begin{eqnarray}
\delta Q& \equiv &\langle
A_{n-2}\stackrel{m}{\Psi},\xi\rangle=\frac{\kappa}{8\pi
G_n}\langle dA_{n-2},\xi \rangle-\langle
A_{n-2}\stackrel{e}{\Psi},\xi\rangle, \nonumber \\
&=&\frac{\kappa}{8\pi G_n}\langle dA_{n-2},\xi
\rangle+\frac{n-2}{n-1}\frac{\kappa_{n+1}^4}{32\pi
G_{n}}(1-\epsilon)\rho(\rho+p)A_{n-2}H\tilde{r}_A\, .
\end{eqnarray}
Using the  Friedmann equations given in the above, we arrive at
\begin{eqnarray}
\delta Q &=&-\frac{\epsilon(1-\epsilon)}{4\pi
G_n}(n-2)A_{n-2}H\tilde{r}_A\left(\frac{1}{\sqrt{\tilde{r}_A^4+\ell^2\tilde{r}_A^2}}\right)\nonumber\\
&=&T\langle \frac{ n-2
}{4G_{n}}\frac{\Omega_{n-2}\tilde{r}_A^{n-2}}{\sqrt{\tilde{r}_A^2+\ell^2}}d\tilde{r}_A,
\xi \rangle=T\langle dS,\xi \rangle=TdS\, ,
\end{eqnarray}
where $T=\kappa/2\pi$ and
\begin{equation}
\label{2eq28}
 S=\frac{(n-2)\Omega_{n-2}}{4G_n}{\displaystyle\int^{\tilde
r_A}_0\frac{\tilde{r}_A^{n-2}
}{\sqrt{\tilde{r}_A^2+\ell^2}}d\tilde{r}_A}\,.
\end{equation}
Integrating (\ref{2eq28}), we obtain the entropy expression
associated with the apparent horizon
\begin{equation}
\label{2eq29}
S=\frac{A_{n-2}}{4G_n}\cdot\left\{\frac{n-2}{n-1}\left(\frac{\tilde{r}_A}{\ell}\right)
{}_2F_1\left[\frac{n-1}{2},\frac{1}{2},\frac{n+1}{2},
-\left(\frac{\tilde{r}_A}{\ell}\right)^2\right]\right\}\, ,
\end{equation}
where ${}_2F_1[\alpha,\beta,\gamma,z]$ is Gaussian hypergeometric
function. This entropy expression looks a bit complicated, but has
the expected properties: in the large horizon limit, it reduces to
the standard area formula in $n$ dimensions, while in the small
horizon limit, it becomes the area formula in the bulk. Further,
from the point of view of bulk, one can also arrive at the entropy
expression~\cite{CC2}.

 We can rewrite the unified first law as
\begin{equation}
\label{2eq30}
d\stackrel{m}{E}=A_{n-2}\stackrel{m}{\Psi}+\stackrel{m}{W}dV_{n-2},
\end{equation}
where $\stackrel{m}{E}=\rho V_{n-2}$. After projecting along the
horizon using $\xi$, equation (\ref{2eq30}) gives us with
\begin{equation}
\langle d\stackrel{m}{E},\xi\rangle=\langle TdS,
\xi\rangle+\langle\stackrel{m}{W}dV_{n-2},\xi\rangle.
\end{equation}
This relation is nothing but the first law of thermodynamics
associated with the apparent horizon, $dE=TdS+WdV_{n-2}$, with
identifying the interior energy $E$ to be $\rho V_{n-2}$,
temperature $T$ to $\kappa/2\pi$, entropy $S$ to the form given
in~(\ref{2eq29}), and the work density $W=(\rho-p)/2$.

As a result, it turns out that we can indeed obtain an entropy
expression for the apparent horizon in the RS model by applying
the method proposed in \cite{CC1}, and the universal relation,
$dE=TdS+WdV_{n-2}$, also holds for the apparent horizon in this
brane world scenario.

\section{Entropy of apparent horizon in warped DPG model}

The original DGP model consists of a bulk scalar curvature term
and an intrinsic scalar curvature term on the brane without
tension~\cite{DGP}. Here we consider the so-called warped DGP
model, which essentially is a generalization of the RS II model by
introducing an intrinsic curvature term on the brane. Considering
an $(n-1)$-dimensional brane with perfect fluid in the warped DGP
model, we have the corresponding Friedmann equation~\cite{SWC1}
\begin{equation}\label{3eq1}
\sqrt{H^2+\frac{k}{a^2}-\frac{2\kappa_{n+1}^2\Lambda_{n+1}}{n(n-1)}-\frac{\mathcal{C}}{a^n}}
=-\frac{\kappa_{n+1}^2}{4\kappa_{n}^2}(n-2)(H^2+\frac{k}{a^2})+\frac{\kappa_{n+1}^2}{2(n-1)}\rho,
\end{equation}
where $H=\dot{a}/a$ is the Hubble parameter on the brane and
${\cal C}$ denotes for the dark radiation term. In the following
we set ${\cal C}=0$. And the energy density on the brane obeys the
continuity equation, $\dot \rho +(n-1)H(\rho+p)=0$. In this
section, we derive the entropy expression of the apparent horizon
on the brane by using the relation
 \begin{equation}
 \label{3eq2}
 dE=TdS +WdV,
 \end{equation}
 where the definitions of these quantities are the same as those
 in the previous section.

The apparent horizon radius in the FRW universe is $ \tilde r_A=
1/\sqrt{H^2+k/a^2}$, and the surface gravity has the form $ \kappa
= -\frac{1}{\tilde{r}_A} \left( 1-\frac{\dot{\tilde r}_A}{2H\tilde
r_A}\right).$ The temperature associated with the apparent horizon
is defined by $T=\kappa/2\pi$.

1) Let us first consider the case without the intrinsic curvature
term, namely $\kappa_n^2 \to \infty $ in (\ref{3eq1}).  This case
can be divided into two subcases. (i) The brane is embedded into a
Minkowskian space, namely, the bulk cosmological constant
$\Lambda_{n+1}$ vanishes. In this case, we find that by using the
relation $dE=TdS +WdV$, the entropy has the form
\begin{eqnarray}\label{3eq3}
S = {\displaystyle \int^{\tilde r_A}_0 dS }
&=&\frac{(n-1)\Omega_{n-1}}{2G_{n+1}}{\displaystyle \int^{\tilde
r_A}_0 \tilde {r}_{A}^{n-2}d\tilde
{r}_{A}}=\frac{2\Omega_{n-1}\tilde {r}_{A}^{n-1}}{4G_{n+1}}.
\end{eqnarray}
This entropy obeys the area formula in $(n+1)$ dimensions. The
factor $2$ comes from the $Z_2$ symmetry of the brane. This is an
expected result since in this case, the gravity on the brane is
the $(n+1)$-dimensional Einstein gravity, the localization of
gravity does not happen.

 (ii) The brane is embedded in an AdS
space. This case is just the one discussed in the previous
section. Indeed, we can obtain the entropy expression like
(\ref{2eq29}).

2) The case with the intrinsic curvature term. This case can also
be divided into two subcases. (i) The case with the brane being
embedded in a Minkowskian space.  This case through $dE=TdS +WdV$,
we obtain
\begin{eqnarray}\label{3eq4}
 S &=&
(n-1)\Omega_{n-1}{\displaystyle\int^{\tilde
r_A}_0\left(\frac{(n-2){\tilde{r}_A}^{n-3}}{4G_n}+\frac{{\tilde{r}_A}^{n-2}}{2G_{n+1}}
\right)d\tilde{r}_A} \nonumber \\
&=&\frac{(n-1)\Omega_{n-1}{\tilde{r}_A}^{n-2}}{4G_n}+\frac{2\Omega_{n-1}{\tilde{r}_A}^{n-1}}{4G_{n+1}}=S_{n}+S_{n+1}.
\end{eqnarray}
It is interesting to note that in this case the entropy can be
regarded as a sum of two area formulas; one (the first term)
corresponds to the gravity on the brane and  the other (the second
term) to the gravity in the bulk. This indeed reflects the fact
that there are two gravity terms in the action of DGP model.

(ii) The brane is embedded in an AdS space. This time we arrive at
\begin{equation} \label{3eq5}
S=\frac{(n-1)\Omega_{n-1}{\tilde{r}_A}^{n-2}}{4G_{n}}+\frac{2\Omega_{n-1}{\tilde{r}_A}^{n-1}}{4G_{n+1}
}
 \times
{}_2F_1\left(\frac{n-1}{2},\frac{1}{2},\frac{n+1}{2},
-\frac{{\tilde{r}_A}^2}{\ell^2}\right).
\end{equation}
This is also a sum of two terms. This first term is nothing, but
the are formula coming from the intrinsic curvature term on the
brane, while the second term is just the one we obtained in the
RSII model.

\section{Entropy of apparent horizon in the Gauss-Bonnet brane world with induced gravity }

The approach in the previous section can be applied to a more
general case to obtain an entropy expression associated with
apparent horizon in brane world scenarios. In this section we
consider a general case, where both an intrinsic curvature term on
the brane and a scalar curvature term in the bulk appear, besides,
a Gauss-Bonnet term are also present in the bulk~\cite{SWC2}. The
action of the model under consideration is
\begin{equation}
\label{4eq1} S  =  \frac{1}{2{\kappa_5}^2} \int{
d^5x\sqrt{-{g}}\left({R}-2\Lambda+\alpha \mathcal{L}_{GB}\right)}
  + \frac{1}{2{\kappa_4}^2}\int
{d^4x\sqrt{-\widetilde{g}}\widetilde{R}}+\int
{d^{4}x\sqrt{-\widetilde{g}}( \mathcal {L}_{m} -2 \lambda)},
\end{equation}
where $\Lambda<0$ is the bulk cosmological constant and ${\mathcal
L}_{GB}=R^2-4R^{AB}R_{AB}+R^{ABCD}R_{ABCD}$ is the Gauss-Bonnet
correction term. In this case, the corresponding Friedmann
equation on the brane is~\cite{Kon,SWC2}
\begin{equation}\label{4eq2}
\epsilon\frac{2{\kappa_4}^2}{{\kappa_5}^2}\left[1
+\frac{8}{3}\alpha\left(H^2 +{k \over a^2} + {\Phi_0\over 2}
\right) \right]\left(H^2 +{k \over a^2}-\Phi_0\right)^{1/2}
=-\frac{\kappa^{2}_{4}} {3}\rho+H^2 +{k \over a^2} ,
 \end{equation}
where $\Phi_0=
\frac{1}{4\alpha}\left(-1+\sqrt{1-\frac{8\alpha}{\ell^2}}\right)$
 and $\epsilon=\pm1$. For later convenience we choose $\epsilon=-1$.
 Furthermore, here we have set the bulk dark radiation to vanish.

 In this case, we find
 \begin{equation}
 \label{4eq3}
dE -WdV = T \left(\frac{3\Omega_{3}}{2G_{4}}\tilde{r}_{A}
+\frac{3\Omega_{3}}{2G_{5}}\frac{\tilde{r}_{A}^{2}}{\sqrt{1-\Phi_0
\tilde {r}_{A}^2}}\right)d\tilde {r}_{A} + T
\frac{6\alpha\Omega_{3}}{G_{5}}\left(\frac{2-\Phi_0 \tilde
{r}_{A}^2 }{\sqrt{1-\Phi_0 \tilde {r}_{A}^2}}\right)d\tilde
{r}_{A}.
\end{equation}
Thus the entropy associated with the apparent horizon can be
obtained as
 \begin{equation} \label{4eq4}
 S = \frac{3\Omega_{3}}{2G_{4}}{\displaystyle\int^{\tilde
r_A}_0\tilde{r}_{A}d\tilde {r}_{A}}+
\frac{3\Omega_{3}}{2G_{5}}{\displaystyle\int^{\tilde r_A}_0
\frac{\tilde{r}_{A}^{2}d\tilde {r}_{A}}{\sqrt{1-\Phi_0 \tilde
{r}_{A}^2}}} +
\frac{6\alpha\Omega_{3}}{G_{5}}{\displaystyle\int^{\tilde r_A}_0
\frac{2-\Phi_0 \tilde {r}_{A}^2 }{\sqrt{1-\Phi_0 \tilde
{r}_{A}^2}}d\tilde {r}_{A}}.
\end{equation}
Integrating this yields an explicit form of the entropy
\begin{eqnarray} \label{4eq5}
S
 &=&\frac{3\Omega_{3}{\tilde{r}_A}^{2}}{4G_{4}}+\frac{2\Omega_{3}{\tilde{r}_A}^{3}}{4G_{5}}
{}_2F_1\left(\frac{3}{2},\frac{1}{2},\frac{5}{2},
\Phi_0{\tilde{r}_A}^2\right)
 \nonumber \\
 && +\frac{6\alpha\Omega_{3}{\tilde{r}_A}^3}{G_{5}} \left(
{}_2F_1\left(\frac{3}{2},\frac{1}{2},\frac{5}{2},
\Phi_0{\tilde{r}_A}^2\right) \Phi_0 +\frac{\sqrt{1-\Phi_0 \tilde
{r}_{A}^2}}{{\tilde{r}_A}^2}\right),
\end{eqnarray}
where ${}_2F_1(a,b,c,z)$ is a hypergeometric function. The
expression looks complicated. But its physical meaning is clear
and includes various special cases~\cite{SWC2}. In particular,
keeping $\alpha$ finite, and taking the limit $G_4 \rightarrow
\infty $ and  $\Phi_0 \rightarrow 0$, one can extract from Eq.
(\ref{4eq5}) the entropy associated with the apparent horizon on
the brane in the Gauss-Bonnet brane world with a Minkowskian bulk
$ S=\frac{2\Omega_{3}{\tilde{r}_A}^{3}}{4G_{5}}
 \left(1+\frac{12 \alpha}{{\tilde{r}_A}^{2}}\right).
$ This expression  is the same as that of entropy of black holes
in the Gauss-Bonnet gravity. This give a self-consistency check
for the entropy (\ref{4eq5}).

\section*{Acknowledgements}
I would like to thank my collaborators, M. Akar, L.M. Cao, S.P.
Kim, A. Sheykhi and B. Wang for collaboration in these topics
discussed in this work, and also the organizers for inviting me to
take part and to present this work in the wonderful conference,
ICGA8. This work was supported in part by a grant from Chinese
Academy of Sciences, by NSFC under grants No. 10325525 and No.
90403029.

%

\end{document}